\documentclass[aps,pra,twocolumn,unsortedaddress,superscriptaddress,showkeys]{revtex4-1}
\usepackage{multirow}
\usepackage{color}
\usepackage{graphicx}
\usepackage{natbib}
\usepackage{amsmath}
\begin{document}


\title{Photoresponse dynamics in amorphous-LaAlO$_{3}$/SrTiO$_{3}$ interfaces}
\author{Emiliano Di Gennaro}
\email[Corresponding author: ]{emiliano@na.infn.it}
\affiliation{Dipartimento di Fisica, Univ. di Napoli Federico II and CNR-SPIN, Compl. Univ. di Monte S. Angelo, Via Cinthia I-80126 Napoli (Italy)}
\author{Ubaldo Coscia}
\affiliation{Dipartimento di Fisica, Univ. di Napoli Federico II and CNISM Unit\'a di Napoli, Compl. Univ. di Monte S. Angelo, Via Cinthia I-80126 Napoli (Italy)}
\author{Giuseppina Ambrosone}
\affiliation{Dipartimento di Fisica, Univ. di Napoli Federico II and CNR-SPIN, Compl. Univ. di Monte S. Angelo, Via Cinthia I-80126 Napoli (Italy)}
\author{Amit Khare}
\affiliation{Dipartimento di Fisica, Univ. di Napoli Federico II and CNR-SPIN, Compl. Univ. di Monte S. Angelo, Via Cinthia I-80126 Napoli (Italy)}
\author{Fabio Miletto Granozio}
\affiliation{Dipartimento di Fisica, Univ. di Napoli Federico II and CNR-SPIN, Compl. Univ. di Monte S. Angelo, Via Cinthia I-80126 Napoli (Italy)}
\author{Umberto Scotti di Uccio}
\affiliation{Dipartimento di Fisica, Univ. di Napoli Federico II and CNR-SPIN, Compl. Univ. di Monte S. Angelo, Via Cinthia I-80126 Napoli (Italy)}

\date{\today}

\begin{abstract}
The time-resolved photoconductance of amorphous and crystalline LaAlO$_3$/SrTiO$_3$ interfaces, both hosting an interfacial 2-dimensional electron gas, is investigated under irradiation by variable-wavelengths, visible or ultraviolet photons. Unlike bare SrTiO$_3$ single crystals, showing relatively small photoconductance effects, both kinds of interfaces exhibit an intense and  highly persistent photoconductance with extraordinarily long characteristic times. The temporal behaviour of the extra photoinduced conductance persisting after light irradiation shows a complex dependence on interface type (whether amorphous or crystalline), sample history and irradiation wavelength. \textcolor{black}{The experimental results indicate that different mechanisms of photoexcitation are responsible for the photoconductance of crystalline and amorphous  LaAlO$_3$/SrTiO$_3$ interfaces under visible light. We propose that the response of crystalline samples is mainly due to the promotion of electrons from the valence bands of both SrTiO$_3$ and LaAlO$_3$. This second channel is less relevant in amorphous LaAlO$_3$/SrTiO$_3$, where the higher density of point defects plays instead a major role.}
\end{abstract}
\pacs{}
\keywords{Interfaces, 2D Electron Gas, Photoconductivity, Oxides}
\maketitle
%
%
The interface between the band insulators LaAlO$_3$(LAO) and SrTiO$_3$(STO) can host a 2-dimensional electron gas (2DEG). In the seminal paper by \textcolor{black}{Ohtomo and Hwang,\cite{Ohtomo2004}} an epitaxial LAO film was grown on a single crystal (001) STO substrate with single TiO$_2$ termination. In this heterostructure, the alternating  LaO and AlO$_2$ planes carry opposite charges, leading to a net polarization of the LAO film, in contrast with the non-polar state of STO. Such polar discontinuity was indicated as a source of instability, resulting in the injection of electrons from the topmost LAO layers into the Ti3d states at the interface, thus giving rise to the interfacial 2DEG. \cite{Ohtomo2004} The above described "electronic reconstruction" (ER) model is confirmed by ab initio computations\cite{Popovic2008,Pentcheva2009, Stengel2011} and provides a natural explanation of some well established experimental results, as the need of TiO$_2$-terminated STO substrates and a LAO film critical thickness of 4 unit cells to achieve interface conductivity.\cite{Nishimura2004,Thiel2006} 
Experiments and theoretical calculations also agree on the bending of the electronic bands of STO close to the interface, as an effect of the local electric field. Such bending determines a triangular quantum well that confines the 2DEG within a few nanometers. \cite{Basletic2008, Stengel2011, Cantoni2012}

It was soon realized that, beside the ER model, other mechanisms may explain the presence of free electrons populating of the quantum well. In particular, each oxygen vacancy introduced in STO acts as an electron donor, thus explaining the observed dependence of the carrier number on the oxygen partial pressure during the LAO film growth.\cite{Huijben2009,Cancellieri2010} It was also suggested that the oxygen vacancies are formed close to the interface, during sample fabrication, because the oxidation of LAO partially occurs by taking O atoms from the STO substrate. \cite{Schneider2010} Furthermore, oxygen vacancies can produce a 2DEG even when an ER cannot take place, as for instance in  the interface between oxygen-reduced STO and vacuum.\cite{Santander2011,DiCapua2012} A spectacular demonstration of such behavior in LAO/STO interfaces was provided by the work of Chen, et al., \cite{Chen2011} showing that the interface between crystalline STO and amorphous LAO can be conductive. Recently, the role of oxygen vacancies in interfaces based on either a crystalline (c), or an amorphous (a) LAO layers on STO was further clarified by observing the effects of thermal and ion irradiation treatments on both kinds of samples.\cite{Liu2013,Rubano2014}  The results suggest that in c-LAO/STO the conductivity can be ascribed to both ER and oxygen vacancies, depending on the sample growth conditions, whereas in a-LAO/STO only oxygen vacancies play a role.

The sensitivity of c-LAO/STO to the light was discovered several years ago.\cite{Huijben2006a, Huijben2006b, Thiel2009} More recently, the investigation of the transport properties of irradiated samples revealed its potential interest for both applicative and fundamental physics.\cite{Irvin2010, Rastogi2012b, Tebano2012, DiGennaro2013, Lu2013,Lu2014,Liang2013} Photoconductivity was reported to be moderate when samples were illuminated by visible light, and it became considerably larger above a photon energy threshold of about 3.25 eV (382 nm),\cite{Irvin2010} corresponding to the indirect bandgap of STO. A striking feature of c-LAO/STO is the persistence of the photocurrents after restoring the dark conditions; the characteristic times range from 10 to 10$^4$ seconds or even more, depending on the sample. In presence of point defects, bare STO can also show persistent photoconductivity,\cite{Tarun2013} resembling the classical lattice-relaxation of doped GaAs, CdZn, CdS, etc.\cite{Lang1977} The case of c-LAO/STO may instead be associated with, e.g., GaAs/AlGaAs heterostructures, where the persistent photoconductivity is ascribed to the spatial separation of the electron-hole pair under the effect of the local electric field.\cite{Queisser1986, Kioupakis2012}

In this paper we investigate the photoresponse of both a- and c-LAO/STO. Our aim is to provide a wider characterization of the interfaces based on amorphous LAO, that are certainly less explored, and to compare their behavior to that of crystalline interfaces. Potentially, this is a very interesting issue, because the crucial difference between the two systems consists in the amount of oxygen vacancies that, acting as point defects, may play a key role in the photoresponse. 
\textcolor{black}{In the following, we present data regarding two samples of a-LAO/STO with different thickness, one c-LAO/STO and one bare STO substrate. We stress that, according to the mentioned attitude of LAO to oxidize at the expense of STO, and due to the different growth conditions we adopted, the investigated a-LAO/STO samples contain much more point defects, i.e., oxygen vacancies, than c-LAO/STO. Further details on the fabrication are given in the Method section}. It turned out that the time evolution of the photoconductance, during the exposition to radiation of different wavelengths in the visible-ultraviolet (VIS-UV) region, and after having turned the light irradiation off, shows clear differences between different types of samples. \textcolor{black}{We analyze the possible microscopic mechanisms of carrier photogeneration, and propose that the different behavior is ascribed to different band diagrams for both structures.  This is discussed in terms of naive diagrams for a- and c-LAO/STO.}

\section*{Results and Discussion}
The four samples used in transport and photo-transport measurements are labelled as follows:\\
\textbf{C1} - a conducting 6 u.c. (thickness d$\simeq$2.4 nm) c-LAO/STO heterostructure;\\
\textbf{A1} - a conducting a-LAO/STO interface, with an estimated LAO thickness d$\simeq$2.4 nm (similar to C1);\\
\textbf{A2} - an insulating a-LAO/STO interface, with an estimated LAO thickness d$\simeq$1.2 nm;\\
\textbf{S1} - a bare TiO$_2$-terminated STO single crystal. 

In dark, S1 exhibits a very high resistance (250 T$\Omega$ at room temperature, close to the limit of our measurement setup), characteristic of fully oxidized, non doped STO crystals. The deposition of a very thin amorphous LAO overlayer (sample A2) determines a noticeable change: the room temperature resistance in dark is lower by 4 orders of magnitude with respect to S1 (Table  \ref{tbl:expdata}). 

\begin{table}[h]
\small
\caption{Transport data at room temperature in dark condition for all samples.}
  \label{tbl:expdata}
  \begin{tabular*}{0.45\textwidth}{@{\extracolsep{\fill}}|c|cccc|}
    \hline
     Sample &d (nm) &V (V) &I$_d$ (A)&R$_d$ ($\Omega$)\\
     \hline
   C1&  2.4&    1&      23.2 $\mu$&     43 k\\
A1&     2.4&1&  15.6 $\mu$&     64 k\\
A2&     1.2&    30&     3.5 n&  8.6 G\\ 
S1&     -&      500&2 p&        250 T \\
    \hline
  \end{tabular*}
\end{table}

\begin{figure}[bt!]
\centering
  \includegraphics[width=0.45\textwidth]{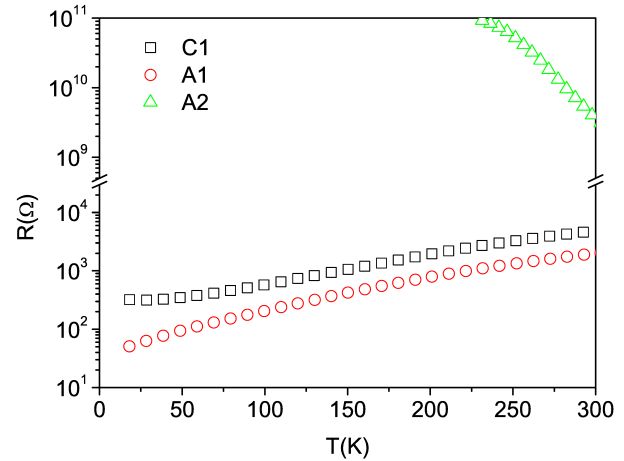}
  \caption{Resistance vs. temperature plots of C1, A1 and A2 samples}
  \label{fig1}
\end{figure}

However, A2 still shows an insulating behavior, i.e., a negative slope in the R(T) plot (fig. \ref{fig1}), suggesting that the fabrication process introduced some oxygen vacancies in the structure, but their amount is still insufficient to determine the formation of a mobile 2DEG. When the amorphous LAO film is thicker (A1), a dramatic drop of resistance and a metallic behavior (positive slope in the R(T) plot) are observed, in agreement with literature.\cite{Chen2011, Liu2013} Basically, A1 shares with the crystalline interface C1 quite similar transport properties in dark.

\begin{figure}[th]
\centering
  \includegraphics[width=0.45\textwidth]{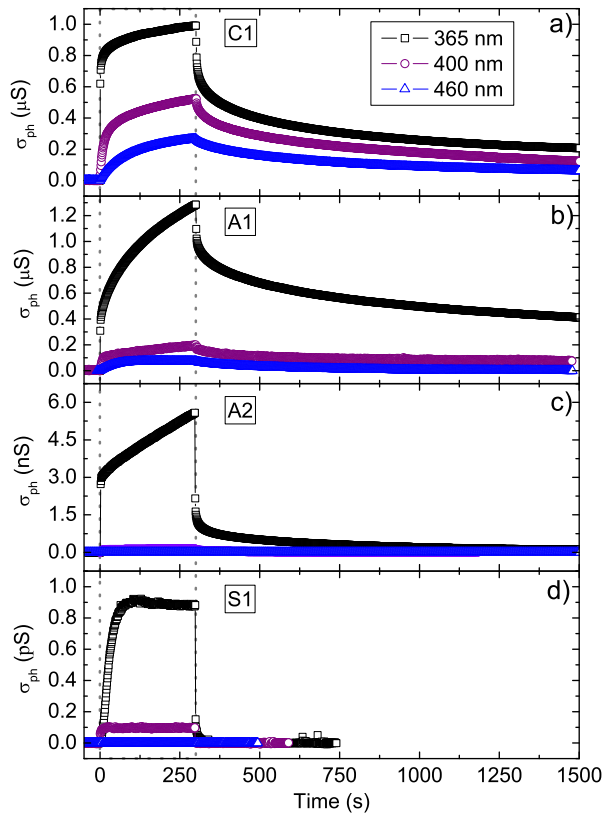}
  \caption{a), b), and c) Photoconductance vs. time of C1, A1 and A2 interfaces; d) photoconductance vs. time of bare STO (sample S1). The photon energy at 365 nm (3.4 eV) is  above the indirect STO gap; at 400 nm (3.1 eV) and 460 nm (2.7 eV) it is below this threshold. All samples were illuminated for the same time (for 300 s) at equal photon flux ($\sim$ 2x10$^{14}$ photons cm$^{-2}$s$^{-1}$) for each wavelength.} 
  \label{fig2}
\end{figure}

To measure the samples photoresponse we select three different wavelengths: 365  nm (3.4  eV, above the indirect badgap of STO); 400 nm (3.1 eV, below the indirect badgap of STO); 460 nm (2.7 eV, well below the indirect badgap of STO).
We define the photocurrent I$_{ph}$(t) as the difference between the current I(t), measured at time t during the light illumination, and the current in dark conditions before the irradiation, I$_d$. Then, the photoconductance is given by  $\sigma_{ph}(t)=\frac{I_{ph}(t)}{V}$, V being the voltage drop across the two probes. The time dependent photoconductance of C1, A1, A2, and S1 at three different wavelengths is reported in fig. \ref{fig2}. 

\textcolor{black}{The photoresponse dynamics of all the measured samples show a relatively steep increase, followed by a slower transient behavior}. However, the intensity is dramatically different for each case. In table \ref{tbl:lightdata} we report the value of photoconductance after 300 s of illumination, $\sigma_{ph}^{max}$ , for each sample and wavelength. At 365 nm, $\sigma_{ph}^{max}$ is of the order of $\mu$S for A1 and C1 samples, while it is 3 and 6 orders of magnitude smaller for A2  and S1, respectively. Conversely, the photosensitivity $\Psi_{ph}(t)=\frac{\sigma_{ph}(t)}{\sigma_{d}}=\frac{I_{ph}(t)}{I_{d}}$  is maximum in S1, lower in A2 and even lower in A1 and C1.

These preliminary observations allow to outline  the general framework. A set of donor states (possibly different for S1, A1, A2 and C1) can be excited by photons. The photoresponse is directly related to the number $\Delta$n of electrons promoted into the conduction band. In the steady-state regime, $\Delta$n is determined by the balance between the excitation and the recombination rates. The enhancement of the photoconductance in the sequence S1, A2, and C1/A1 then suggests that the recombination rate becomes lower and lower when the LAO thickness increases. Under this point of view, our results allow to extend to the case of a-LAO/STO a previous interpretation of the c-LAO/STO photoresponse.\cite{DiGennaro2013} The argument is also qualitatively confirmed by the different duration of the tails after turning  the illumination off; in fact, the recovery rate increases in the sequence C1/A1, A2, S1. 

However, this enhancement may also have an alternative explanation. It has been proposed \cite{Popovic2008} that the deepest electronic states in the quantum well of c-LAO/STO interfaces have low or null mobility. It is reasonable to assume that the same also holds for a-LAO/STO, since the electronic properties of the 2DEG are mainly determined by the STO crystal in both cases. In dark conditions, these states are completely filled in the conducting interfaces, so that all the electrons promoted by the light must contribute to the conduction. Instead, they are (at least partially) empty in A2. Therefore, a fraction of the photo-generated electrons may occupy these states and may not contribute to the electrical conduction of A2, i.e., it  may not be detected in the photoresponse measurements.

Since the same photon flux is adopted at all wavelength, and considering that each process of photo-excitation generates one free carrier, it follows that the variation of photoconductance with wavelength just reflects the variation of photogeneration efficiency. The observation that the maximum photocondutance $\sigma_{ph}^{max}$ systematically increases at decreasing the radiation wavelength is then easily understood, as higher energy photons can excite deeper states; and, in particular, photons at 365 nm can, in principle, promote electrons from the STO valence band (VB) to the conduction band (CB). 

The ratio between the values of $\sigma_{ph}^{max}$ at 365nm and 460 nm is higher in the insulating samples (S1, A2), than in the conducting samples (A1, C1). Still in the same spirit of a na\"{i}ve interpretation, this suggests that the trap states form a narrow band, close to the valence band edge, in S1 and in A2, so that low energy photons can hardly promote electrons to the CB. The trap states form instead a broad band in A1, and still broader in C1. In other terms, the effective optical gap at the interface is always smaller than the band gap of bulk STO; and it is smaller in c-LAO/STO interfaces than in a-LAO/STO ones.
\begin{table}[h] 
\small
\caption{Photoconductance and photosensitivity values after 300 s of illumination for the three different  wavelengths.}
  \label{tbl:lightdata}
  \begin{tabular*}{0.35\textwidth}{@{\extracolsep{\fill}}|c|ccc|}
  \hline
Sample&$\lambda$(nm)&$\sigma_{ph}^{max}$ (nS)&$\Psi^{max}$\\

\hline 
\multirow{3}{*}{C1}& 365& 991&0.043\\
						& 400	& 524& 0.023\\
						&460&267&	0.012\\
     \hline
\multirow{3}{*}{A1}& 365& 1285&0.082\\
						& 400	& 285& 0.018\\
						&460&75&	0.005\\
	\hline
\multirow{3}{*}{A2}& 365& 6.3&54\\
						& 400	& 0.666& 5.7\\
						&460&0.296&	2.5\\
	\hline
\multirow{3}{*}{S1}& 365& 8.8$\cdot10^{-4}$ &220\\
						& 400	& 9.8$\cdot10^{-5}$ &25\\
						&460& 1.2$\cdot10^{-5}$ &3\\
\hline
  \end{tabular*}
\end{table}

In order to get a more accurate description of the photoresponse, it is necessary to discuss the dynamic behavior of the signals. We first start addressing the issue of the increase of photoconductance as a function of the illumination time (see fig. \ref{fig2}). The plots of $\sigma_{ph}$(t) are characterized by a steep onset followed by a slower transient. This behavior is well described by a bi-exponential growth function with two characteristic rise times $\tau_1$ and $\tau_2$, related to the fast and slow components respectively:
\begin{align}
\label{eq1}
  \sigma_{ph}(t)=\sigma_{inf}\left\lbrace A\left[ 1-exp\left(-\frac{t}{\tau_{1}}\right)\right]+ \right. \nonumber \\ 
  \left. (1-A) \left[ 1-exp\left(-\frac{t}{\tau_{2}}\right)\right]\right \rbrace
\end{align}
where $\sigma_{inf}$ is the saturation value of the photoconductance and A represents the relative weight of the fast component. 
\begin{figure}[th]
\centering
  \includegraphics[width=0.45\textwidth]{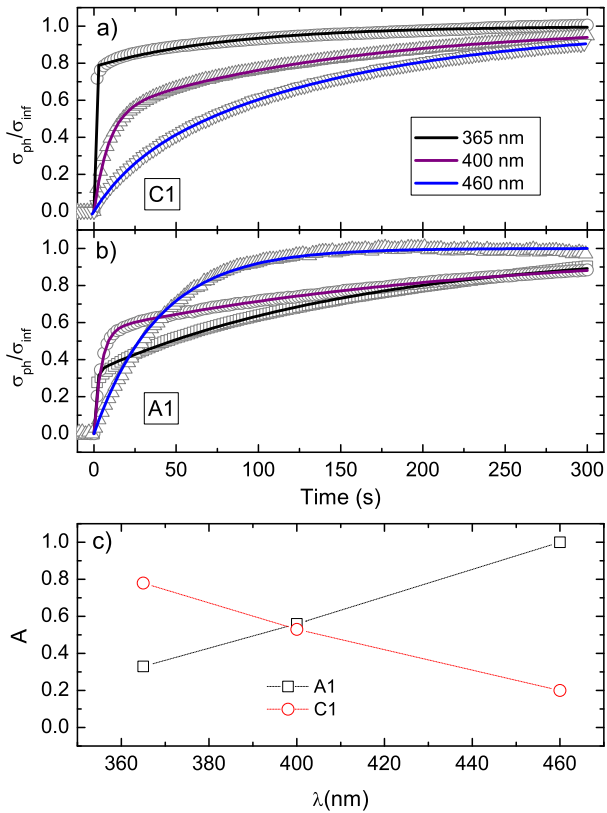}
  \caption{a) Photoresponse of C1 and b) photoresponse of A1 at 365 nm (black), 400 nm (red), and 460 nm (green). The data are normalized to the asymptotic value $\sigma_{inf}$; solid lines are fit curves.  c) Dependence of A vs. radiation wavelength $\lambda$  for both samples.}
  \label{fig3}
\end{figure}

We focus now the attention on the metallic samples, C1 and A1 samples. The fitting parameters are reported in Table \ref{tbl:fitdata}, and the fits are shown in fig. \ref{fig3}. In our experimental conditions, the maximum photoconductance  scales with light intensity, i.e., the optical pumping is below the threshold at which the response falls in the saturation regime. This circumstance allows the connection between the dynamics of the photoconductance and the photogeneration mechanism of carriers. On this basis, we interpret the elongation of the fast time $\tau_1$ at increasing wavelength as an effect of the decrease of photogeneration efficiency. This mechanism is common to both C1 and A1. Instead, a striking difference between the samples is found in terms of the relative amplitude of the fast and slow component of the signal, i.e., the A parameter. This difference is marked by the dependence of A vs. radiation wavelength for the two samples (fig. \ref{fig3}c).
\begin{table}[h]
    \caption{Best fitting parameters for the the three different illumination wavelengths.}
    \small
    \begin{tabular*}{0.45\textwidth}{@{\extracolsep{\fill}}|c|ccccc|}
      \hline
    Sample &$\lambda$(nm) &$\sigma_{inf}$ (nS)& A& $\tau _1$(s)& $\tau _2$(s)\\
    \hline
     \multirow{3}{*}{C1}& 365& 990&     0.78    & 0 & 82\\
                        & 400   & 553& 0.53      & 9    & 147\\
                        &460&295&       0.19& 24&       140\\
     \hline
     \multirow{3}{*}{A1}&365& 1420&      0.33&   1.3&    166\\
						&400    &322&   0.56&   4&      228\\
						&460&74&        1&      40&--\\
      \hline
  \end{tabular*}
  \label{tbl:fitdata}
\end{table}

In order to address this issue, we first remind that the selected wavelengths correspond to photon energies that span from slightly above to well below the indirect STO gap. In sample C1, the fast component is progressively reduced with increasing the wavelength, confirming the previous observations for crystalline samples.\cite{DiGennaro2013} At 460 nm, well below the gap threshold, the fast channel is actually almost suppressed and the signal rise is very slow. The opposite behavior is observed in sample A1, where the relative amplitude of the fast component increases at increasing wavelength. At 460 nm, the slow channel is totally  suppressed, so that the signal is well described by a simple exponential growth with one characteristic time $\tau_1$. Such different response of C1 and A1 undoubtedly indicates a different underlying mechanism of photo-excitation, which distinguishes a- from c-LAO/STO.
 
\begin{figure}[h]
\centering
  \includegraphics[width=0.45\textwidth]{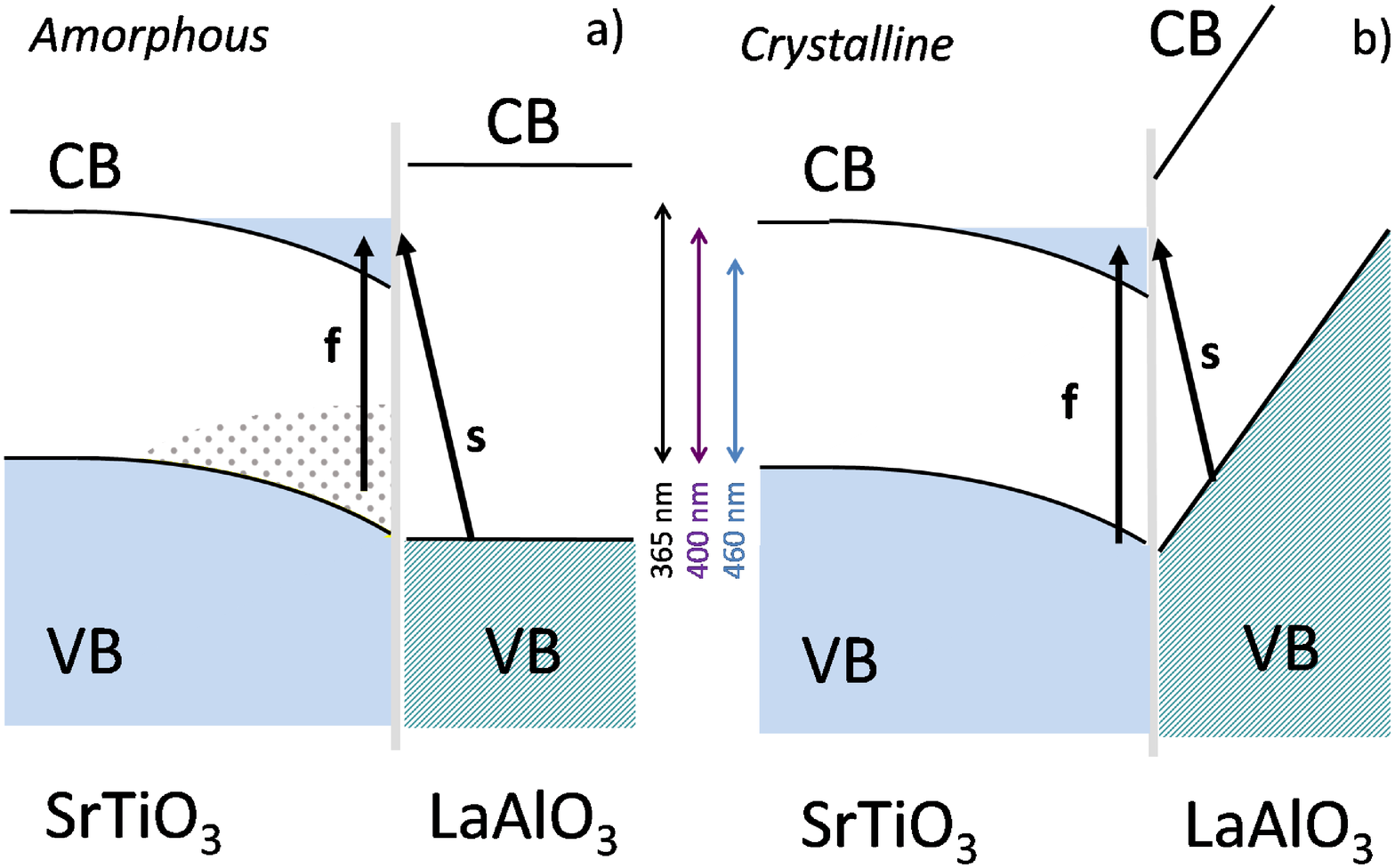}
  \caption{Sketch of the band structure of a) a-LAO/STO and b) c-LAO/STO. The grey dotted region depicts the in-gap \textcolor{black}{trap} states of STO. For clarity, the bending of the STO bands is emphasized in both pictures and, in c-LAO/STO, the band of in-gap states is omitted. The slow (s) and fast (f) mechanisms of photogeneration are represented by arrows.}
  \label{fig4}
\end{figure}

We consider now the delicate issue of the microscopic description of carriers photo-excitation. The modelling of the dynamics in terms of band diagrams is neither straightforward nor unambiguous; however, we believe that the main physics is captured \textcolor{black}{by the naive sketches of fig. \ref{fig4}}. The right panel (fig. \ref{fig4}b) is based on published diagrams for c-LAO/STO; \cite{Rastogi2012a, DiGennaro2013,Liang2013} the left panel (fig. \ref{fig4}a) describes instead a-LAO/STO. The right sides of the band diagrams for a-LAO/STO and c-LAO/STO deeply differ the one from each other and are in agreement with those already reported in \cite{Rubano2014}. For a-LAO/STO, we draw, as an ansatz, a scheme qualitatively similar to the one for bulk LAO crystal. For c-LAO/STO, instead, we adopt the standard ER model, with the bands of LAO bent upwards by the macroscopic electric field. The left part of the diagrams of both a- and c-LAO/STO shows the quantum well that is formed at the interface, within STO, and that confines the mobile electrons in the 2DEG. The point defects, which we mainly identify with oxygen vacancies, add to the bands a distribution of trap states, forming a broad in-gap band. Such states are mainly localized in a region close to the interface, because, as previously mentioned, they are formed during the film growth step.\cite{Liu2013} Due to the different fabrication procedure, the density of vacancies is lower in c-LAO/STO, and as a first approximation we neglect it.
The sketches show some possible channels of photogeneration, labeled as \textit{s} or \textit{f} to indicate slow and fast dynamics, respectively. The main channel of fast excitation is the local (on-site) promotion of electrons from initial states with O-character to final states of the CB with Ti-character; this process may actually involve intermediate states, such as excitons (i.e., localized Ti$^{3+}$ states),\cite{Zhou2011} which rapidly decay into the CB. The slow process is here identified with the promotion of electrons from LAO states with O-character to STO states with Ti-character. These transitions are important, because they can involve initial and final states with non negligible overlap. In the case of c-LAO, for instance, the closest O atom of LAO is actually in the apical site of a Ti atom of the STO side. A similar consideration can be applied also to the case of a-LAO. However, there is a relevant difference between the two cases. All the VB-edge states of LAO share the same energy in a-LAO/STO. Instead, as the sketch in fig. \ref{fig4}b qualitatively shows, the upturn of the bands in c-LAO is so steep, that the VB edge is already lifted by $ \simeq$ 0.2-0.3 eV at the apical sites. The farther the initial states are located, the larger is the energy lift. 

The sketch makes no claim to completeness; more excitation mechanisms can be at play and also be quantitatively relevant (e.g., the transitions from the defect states of STO in c-LAO/STO, \textcolor{black}{such as photoexcitations of carriers from defect-induced deep-level traps as suggested in \cite{Rastogi2014}, or from any other kind of surface trap state).} However, the minimum set of transitions that we show in the diagrams is sufficient to qualitatively explain several experimental observations: 
\begin{enumerate}
\item The reduction of the optical gap in a-LAO/STO (with respect to the indirect STO gap) is explained as an effect of oxygen vacancies; in c-LAO/STO, instead, the reduction of the optical gap is a result of the band bending of LAO. 
\item In c-LAO/STO, the excitation by low energy photons is a slow process, that promotes to the quantum well electrons from the LAO VB; the fast process requires, instead, high energy photons, and consists in the ionization (possibly phonon assisted) of STO VB states.
\item In a-LAO/STO, on the contrary, the excitation by low energy photons is a fast process, promoting to the quantum well electrons from the in-gap trap states and the slow process (promoting electron from the LAO VB) requires high energy photons. Furthermore, the direct excitation of STO VB states in a-LAO/STO may trigger a complex energy cascade within the broad in-gap band. This effect would also indirectly contribute to slow down the dynamics of excitation by high energy photons.
\end{enumerate}

We stress that our data would be hardly explained by assuming that all the relevant transitions are due to the excitation of one type of trap states, i.e., oxygen vacancies only. In particular, this would be at odds with the higher photoconductance induced by low energy photons in c-LAO/STO than in a-LAO/STO (see, e.g., data of $\sigma_{ph}^{max}$ in table \ref{tbl:lightdata}). In fact, the oxygen vacancy states certainly form a broader band in a-LAO/STO than in c-LAO/STO.

\textcolor{black}{We finally add some comments on the bending of c-LAO bands (fig. \ref{fig4}b). In contrast to first-principle computations, \cite{Popovic2008,Pentcheva2009, Stengel2011} the bending was not observed in some x ray photoemission spectroscopy (XPS) measurements.\cite{Chambers2010,Berner2013,Drera2014} In other XPS works (e.g., \cite{Slooten2013}), as well as in cross-sectional scanning tunnelling spectroscopy,\cite{Huang2012} it appeared with reduced slope with respect to expectations. In \cite{Berner2013},  the authors propose two effects (that in our view may concur) to explain the discrepancy between theory and experiment: a) during the XPS experiment, the sample is heavily far from equilibrium due to x ray photoexcitation; b) the surface charge-trap states act as donors, so that even at equilibrium the band bending of c-LAO is attenuated. Both mechanisms may also affect the photoresponse and give quantitatively relevant corrections. However, we believe that the simple qualitative picture that we propose would not change also taking into account these extra channels.}

\begin{figure}[th]
\centering
  \includegraphics[width=0.45\textwidth]{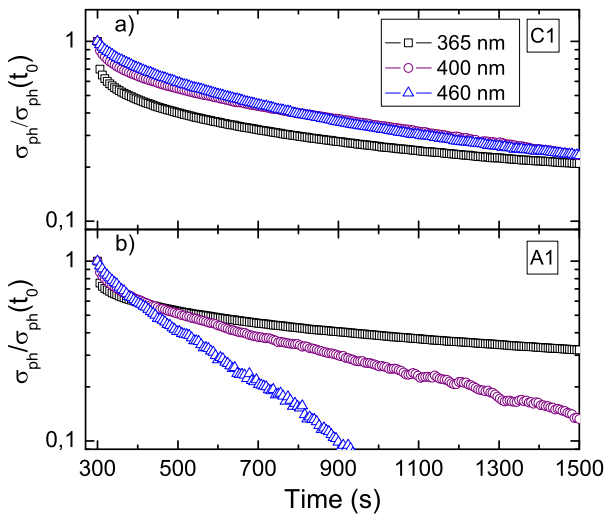}
  \caption{Signal decay after switching off the illumination for a) C1 and b) A1 samples. The same data of fig. \ref{fig2} are normalized to the corresponding $\sigma_{ph}^{max}$ values and plotted in log scale.}
  \label{fig5}
\end{figure}

We turn now to the issue of the signal decay after illumination (fig. \ref{fig5}). As a first comment, we note that this process is affected by the sample history. Depending on the duration and on the intensity of previous expositions to the light, as well as on the wavelength, we observe different values of the final baseline, indicative of metastable states. The further relaxation toward the initial equilibrium state in dark may be extremely long. We did not explore systematically this feature in this work. However it appears that the recovery toward the metastable state is always a process longer than the signal rise one.

As mentioned before, the increase of the characteristic decay time with respect to the rise transient is indicative of a hysteretic behavior that can be attributed to the following mechanisms: a) the activation of a slow energy cascade through intermediate defect states, and b) the drift of photo-generated holes under the action of the macroscopic electric field in the region close to the interface. As previously discussed when commenting the signal rise dynamics, the former mechanism may have comparatively larger weight in a-LAO/STO after UV irradiation. In that case the deep levels in the broad in-gap band are excited, and part of them might be long-lived. The latter mechanism may instead affect the recovery of c-LAO/STO after visible light irradiation, if some holes created in the LAO region are pushed toward the sample surface by the electric field that acts in the polar layer.

\section*{Conclusion}
We investigated the transient photoconductance of amorphous-LAO/STO interfaces, under monochromatic radiation in the UV and Vis region, and compared these systems with a crystalline-LAO/STO structure and a bare STO crystal. In all the samples, the photoconductance assumes the highest value at the shortest explored wavelength (365 nm, corresponding to above-gap excitation). At longer wavelengths, the photoconductance significantly decreases in c-LAO/STO. The reduction is even stronger in conducting a-LAO/STO, and dramatic in the insulating a-LAO/STO sample and in the bare STO crystal. This observation indicates that the optical gap of the interfaces is smaller than the band gap of bulk STO; and that it is the smallest in the case of c-LAO/STO. This effect would be hardly explained by assuming that the photogeneration takes place in all samples by exciting the same states (e.g. in-gap states due to oxygen vacancies), since vacancies certainly generate a broader defects band in a-LAO/STO than in c-LAO/STO.

The transient photoconductance of all samples shows a bi-exponential growth, under illumination, followed by a long decay when the light is turned off. In spite of the qualitative similarities, the photoconductance dynamics reveals a deep difference between a-LAO/STO and c-LAO/STO. In c-LAO/STO, the fast channel of the photoresponse becomes progressively less relevant at increasing photon wavelength; the reverse takes place in a-LAO/STO, where it is the slow channel loosing relative weight at increasing wavelength.

We propose a different mechanism of excitation for a-LAO/STO and c-LAO/STO, which naturally stems from the different band diagrams of the two structures. Indeed, the optical excitation of a-LAO/STO involves transitions from either in-gap trap states due to oxygen vacancies (fast channel) or the LAO VB (slow channel). The first channel is activated by photons in a wide range of energy, while the second is restricted to above-gap photons. The excitation of c-LAO/STO also involves two channels. The first includes the transitions from the STO VB (fast channel) and is restricted to high energy photons; the second includes the transitions from the bent LAO VB band (slow channel), and is activated even by low energy photons.

Finally, a-LAO/STO and c-LAO/STO also have a different behavior of permanent photoconductance when light is turned off, indicating that the physical difference between the structures also determines different dynamics of relaxation.

\section*{Method}
The LAO thin films are grown by a RHEED-assisted Pulsed Laser Deposition technique on TiO$_2$-terminated (001) STO single crystal substrates, resorting to a KrF excimer laser beam with a fluence of 1.5 J/cm$^2$ on target and a repetition rate of 1 Hz. 
The crystalline sample C1 is grown at 720$^\circ$C with a partial pressure of oxygen  PA(O$_2$) = 1x10$^{-2}$mbar. After the deposition, the sample is cooled in 1h to room temperature in the same pressure condition. The amorphous samples A1 and A2 are grown at room temperature with a  PA(O$_2$) = 1x10$^{-5}$mbar.

All the samples are contacted by ultrasonically bonded aluminium wires. The resistance as a function of the temperature R(T) is recorded in dark condition in a closed cycle refrigerator, using the standard four probe Van der Paw configuration for C1 and A1, and a two probe configuration for A2 and S1, due to their high resistance values. In all the photoconductance measurements, the current across the samples is measured at fixed voltage in the two probe configuration, by resorting to a Keithley 6487 picoammeter. All the values of applied voltage V, dark current I$_d$, and resistance in dark R$_d$, are reported in Table \ref{tbl:expdata}.

In order to investigate the time  evolution of the photoconductance, the whole sample surface ($\sim$5$\times$5 mm$^2$) is uniformly illuminated for 300 s with monochromatic radiation obtained by means of a Xenon lamp and of band-pass interference filters (FWHM = 10 $\pm$ 2 nm). Three wavelengths are selected: 365  nm (3.4  eV); 400 nm (3.1 eV); 460 nm (2.7 eV). The power density radiation is measured by a radiometer (Laser Precision RK-5720 Power Radiometer) equipped with a silicon probe. The light intensity was attenuated by means of neutral density filters in order to obtain a photon flux of $\sim$ 2x10$^{14}$ photons cm$^{-2}$s$^{-1}$ for the three wavelengths. 


\section*{Authors contribution}
A.K. and E.D.G. prepared the samples, E.D.G. performed the cryogenic transport measurements, U.C. and G.A. carried out the phototransport measurements, U.S.d.U. and F.M.G. wrote the main manuscript text. All authors reviewed the manuscript. 

\end{document}